\documentclass[epj]{svjour}
\usepackage{graphics}
\begin{document}
\title{Projected shell model study on nuclei near the N = Z line}
%\subtitle{Do you have a subtitle?\\ If so, write it here}
\author{Yang Sun\inst{ }% etc
% \thanks is optional - remove next line if not needed
\thanks{e-mail: ysun@nd.edu}%
}                     % Do not remove
%
%\offprints{}          % Insert a name or remove this line
%
\institute{Department of Physics, University of Notre Dame,
Notre Dame, IN 46556, USA}
\date{Received: date / Revised version: date}
% The correct dates will be entered by Springer
%
\abstract{
Study of the N $\approx$ Z nuclei in the mass-80 region
is not only interesting due to the existence of abundant nuclear structure 
phenomena, 
but also 
important in understanding the nucleosynthesis 
in the rp-process. 
It is not feasible to apply a
conventional shell model due to the necessary 
involvement of the $g_{9/2}$ sub-shell.  
In this paper, the projected shell model is introduced to this study.
Calculations 
are systematically performed for the collective levels 
as well as the quasi-particle excitations. 
It is demonstrated that calculations with this truncation scheme  
can achieve a comparable quality as the large-scale shell model 
diagonalizations for $^{48}$Cr, but the present method can be applied 
to much heavier mass regions. 
While the known experimental data
of the yrast bands in the N $\approx$ Z nuclei (from Se to Ru) 
are reasonably
described, the present 
calculations predict the existence of $K$-isomeric states,  
some of which lie low in energy under certain structure conditions.  
 % end of PACS codes
} %end of abstract
\maketitle
\section{Introduction}
\label{intro}

The proton-rich nuclei with particle number around 80 
exhibit phenomena that are unique to this
mass region. Unlike most heavy nuclei that have a stable deformation, the
structure changes are quite pronounced among the neighboring nuclei, and 
this mass region is
often characterized by shape co-existence. In addition, 
for N $\approx$ Z
nuclei, there is an open question: Whether the neutron-proton correlations
play an important role in the structure analysis. 
Finally, low-lying $K$-isomeric states can
exist in the Z $\sim$ 34 nuclei at an oblate minimum, 
and in the Z $\sim$ 44 nuclei at a prolate one.

The study of these proton-rich nuclei is not only interesting from the nuclear
structure
point of view, but also has important implications in nuclear astrophysics. 
Since heavy elements are made in stellar evolution and explosions,
nuclear physics, and in particular nuclear structure far from stability, enters
into the stellar modelling in a crucial way.
It is believed that these proton-rich nuclei near the N = Z line are synthesised
in the rapid-proton capture process (the rp-process) under appropriate
astrophysical conditions, 
resulting in the creation of nuclei far
beyond $^{56}$Ni all the way to the proton rich regions 
of the chart of nuclides \cite{SchatzRep,SchatzPRL}.
The X-ray burst is suggested as a possible site.
However, the nucleosynthesis and the correlated energy generation  
is not completely understood. In addition to the uncertainty of  
the astrophysical conditions, network simulations of the rp-process  
are hindered by the lack of experimental information on the structure
of nuclei along the rp-process path \cite{SchatzRep}.
Nuclei of particular interest to the rp-process are the N = Z
waiting point nuclei \cite{SchatzRep}.  

Most of the nuclei in the mass-80 region 
are strongly deformed.
At the deformed potential minimum, the high-$j$ $g_{9/2}$ orbital
intrudes into the pf-shell near the Fermi levels.
Therefore, the $g_{9/2}$ orbitals dominate the low-lying 
structure of these nuclei, 
and these orbitals have to be included in the model
space.
A direct inclusion of
the $g_{9/2}$ 
sub-shell into a conventional shell model calculation is not feasible.
This strongly suggests a proper construction of a shell model basis that
is capable of
describing the undergoing physics within a manageable space.

The projected shell model (PSM) \cite{psm} 
is a shell model that builds its model space in a deformed basis. 
Unlike a conventional shell model that starts from a spherical basis,
the PSM uses a deformed basis to start with.  
In this way, 
a large model space can be easily included and  
many nuclear correlations are already constructed before a configuration mixing
is carried out.
This makes shell model calculations for a heavy system possible.
In the long history of the shell model development along this line, 
there have been different types of approaches (see, for example, Refs. 
\cite{psm,MV,FDSM,PSsu3,MC}).  
Although these methods 
differ very much in details in the way of building the shell model bases
and/or choosing effective interactions, they share the common characteristic   
that the model space is first constructed by physical guidance. 
The final diagonalization can therefore be carried out in a highly compact
space, with each of the bases being a complicated combination
in terms of the conventional shell model basis states.

The present paper applies the PSM to the study 
of proton-rich nuclei in the mass-80 region. 
We shall first demonstrate through the example of the $^{48}$Cr calculation 
that the highly truncated calculations performed by the PSM 
can in fact achieve a comparable quality as the large-scale shell model
diagonalizations. However, the present method can certainly be applied
for much heavier mass regions.
Based on the successful reproduction of the known experimental data
of the yrast bands in the N $\approx$ Z nuclei (from Se to Ru), 
the present
calculations further predict the existence of $K$-isomeric states,
some of which lie low in energy under certain structure conditions.
These low-energy isomers could have significant 
effects in the nucleosynthesis
along the rp-process path.
  
\section{The Projected shell model}
\label{sec:1}

\subsection{Outline of the model}
%\label{sec:3}
%as required. Don't forget to give each section
%and subsection a unique label (see Sect.~\ref{sec:1}).

In a PSM calculation, 
the shell-model truncation is first
achieved within the quasi-particle (qp) states with respect
to the deformed Nilsson+BCS vacuum
$\left|\phi\right>$; then rotational
symmetry (and number conservation if necessary) 
are restored for these states by
standard  projection techniques \cite{RS.80} to
form a spherical basis in the laboratory frame; finally the shell model
Hamiltonian is diagonalized in this basis.
If one is mainly interested in the collective rotation with qp
excitation, the 
truncation obtained in this way is very efficient.
Usually, quite satisfactory results can be obtained
by a diagonalization within a dimension smaller than 100.
Clearly, such an approach lies conceptually
between the two extreme methods \cite{fpshell}: deformed mean-field theories and
conventional shell model, and thus can take the advantages
of both.

The central technical question in this regard is how to compute the 
matrix elements in the projected states. Following the pioneering work
of Hara and Iwasaki \cite{HI80}, a systematic derivation has been obtained
for any one- and two-body operators (of separable forces) with an arbitrary
number of quasi-particles in the projected states \cite{psm,code}. 
In principle, the projected multi-qp basis recovers the full shell model
space if all the quasi-particles in the valence space are considered 
in building the multi-qp states.
However, the advantage of working with a deformed basis is that 
the selection of only a few quasi-particles near the Fermi surface 
is already sufficient 
to construct a good shell model space. 
The rest can be simply truncated out.  
 
The set of multi-qp states relevant to the present study 
(of even-even systems) is
\begin{eqnarray}
|\Phi_\kappa\rangle=\{ | 0 \rangle,
\ a^\dagger_{\nu_1}a^\dagger_{\nu_2}| 0\rangle,
\ a^\dagger_{\pi_1}a^\dagger_{\pi_2}| 0\rangle,
\ a^\dagger_{\nu_1}a^\dagger_{\nu_2}a^\dagger_{\pi_1}
a^\dagger_{\pi_2}| 0\rangle \},
\label{conf}
\end{eqnarray}
where $a^\dagger$'s are the qp creation operators, $\nu$'s ($\pi$'s)
denote the neutron (proton) Nilsson quantum numbers which run over
the low-lying orbitals, and $| 0 \rangle$ the
Nilsson+BCS vacuum (or 0-qp state).
Since the axial symmetry is kept for the Nilsson states,
$K$ is a good quantum number. It can be used to label the basis states 
in Eq. (\ref{conf}). 

As in the usual PSM calculations, we use the Hamiltonian of separable forces 
\cite{psm}
\begin{equation}
\hat H = \hat H_0 - {1 \over 2} \chi \sum_\mu \hat Q^\dagger_\mu
\hat Q^{}_\mu - G_M \hat P^\dagger \hat P - G_Q \sum_\mu \hat
P^\dagger_\mu\hat P^{}_\mu,
\label{hamham}
\end{equation}
where $\hat H_0$ is the spherical single-particle Hamiltonian which
in particular contains a proper spin-orbit force, whose strengths (i.e.
the Nilsson parameters $\kappa$ and $\mu$) are taken from Ref.
\cite{brkm}.
The second term in the Hamiltonian is the quadrupole-quadrupole (Q-Q) 
interaction and the last
two terms, the monopole and quadrupole pairing interactions,
respectively. Residual neutron-proton interactions are present only in
the Q-Q term. It was shown \cite{Zuker96} that these interactions
simulate the essence of the most important correlations in nuclei, so
that even the realistic force has to contain at least these components
implicitly in order to work successfully in the structure
calculations. As long as the physics under consideration 
is mainly of the 
quadrupole-and-pairing-type collectivities, we find no compelling reasons 
for not using these simple interactions.
The interaction strengths are determined as follows: the
Q-Q interaction strength $\chi$ is adjusted by the self-consistent
relation such that the input quadrupole deformation $\varepsilon_2$ and
the one resulting from the HFB procedure coincide with each other
\cite{psm}. The monopole pairing strength $G_M$ is taken to be
$G_M=\left[G_1-G_2(N-Z)/A\right]/A$ for neutrons and $G_M=G_1/A$ for
protons, which was first introduced in Ref.~\cite{pairing}. The choice
of the strengths $G_1$ and $G_2$ depends on 
the size of the single-particle space 
in the calculation. 
Finally, the quadrupole pairing
strength $G_Q$ is assumed to be proportional to $G_M$, the
proportionality constant is usually taken in the range of 0.16 --
0.20.

The eigenvalue equation of the PSM for a given spin $I$ takes the
form 
\begin{equation}
\sum_{\kappa'}\left\{H^I_{\kappa\kappa'}-E^IN^I_{\kappa\kappa'}\right\}
F^I_{\kappa'}=0,
\label{psmeq}
\end{equation}
where the Hamiltonian and norm matrix elements are respectively defined
by
\begin{equation}
H^I_{\kappa\kappa'}=\langle\Phi_\kappa|\hat H\hat P^I_{KK'}
|\Phi_{\kappa'}\rangle,~~N^I_{\kappa\kappa'}=\langle\Phi_\kappa|\hat
P^I_{KK'}|\Phi_{\kappa'}\rangle,
\label{elem}
\end{equation}
and $\hat P^I_{MK}$ is the angular momentum projection operator \cite{RS.80}.
The expectation values of the Hamiltonian with respect to a ``rotational
band $\kappa$'' $H^I_{\kappa\kappa}/N^I_{\kappa\kappa}$ are called
the band energies. When they are plotted as functions of spin $I$, we call
it a band diagram \cite{psm}. It usually provides us a useful tool for
interpreting results.

\subsection{An example: $^{48}$Cr}

$^{48}$Cr is a light nucleus for which an exact shell model
diagonalization is feasible, yet exhibits remarkable high-spin phenomena
usually observed in heavy nuclei: large deformation, typical rotational
spectrum and the backbending in which the regular rotational band is
disturbed by a sudden irregularity at a certain spin. This nucleus is
therefore an excellent example for theoretical studies, providing a
unique testing ground for various approaches \cite{fpshell,Cr48}.

\begin{figure}
\resizebox{0.42\textwidth}{!}{%
  \includegraphics{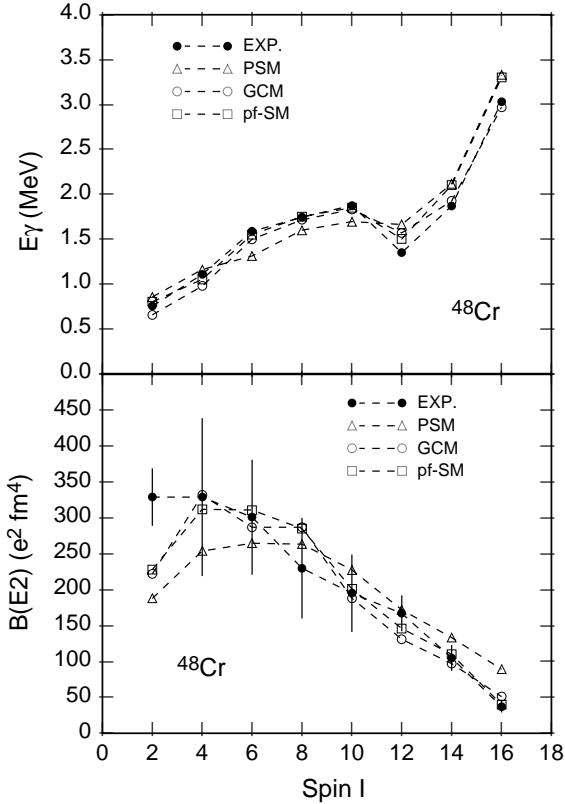}
}
\caption{Top panel: The $\gamma$-ray transition energies $E_\gamma=E(I)-E(I-2)$
as functions of spin. Bottom panel: 
The B(E2) values as functions of spin. The experimental data are taken from Ref.
\protect\cite{exp_b} and the result of pf-SM from Ref.
\protect\cite{fpshell}. The PSM and GCM results 
are taken from Ref. \protect\cite{Cr48}. 
}
\label{fig:0}       % Give a unique label
\end{figure}

In the PSM calculations for $^{48}$Cr, three major shells ($N=1,2,3$) are
used for both neutron and proton. The strengths of the 
monopole pairing forces are chosen as $G_1=22.5$ and $G_2=18.0$ \cite{pairing}.
The quadrupole pairing
strength $G_Q$ is fixed as 0.2$G_M$. The shell model space is truncated at
the deformation  
$\varepsilon_2 = 0.25$. 
 
In the top panel of Fig. 1, the results of the PSM 
for the $\gamma$-ray energy
along the yrast band, together with that of
the pf-shell model (pf-SM)
reported in Ref. \cite{fpshell} and that of the 
Generator Coordinate Method (GCM) \cite{Cr48}, 
are compared with the 
experimental data \cite{exp_b}. One sees that the four curves are
bunched together over the entire spin region, indicating an excellent
agreement of the three calculations with each other, and with the data. The
sudden drop in $E_\gamma$ occurring around spin 10 and 12 corresponds to
the backbending in the yrast band of $^{48}$Cr.

In the bottom panel of Fig. 1, the three theoretical results for B(E2) are 
compared with the
data \cite{exp_b}. All the three calculations use the same effective charges
(0.5e for neutrons and 1.5e for protons). 
Again, one sees that the theoretical descriptions agree not only with
each other but also with the data quite well. The B(E2) values decrease
monotonously after spin 6 (where the first band crossing takes place in
the PSM). This implies a monotonous
decrease of the intrinsic Q-moment as a function of spin, reaching
finally the spherical regime at higher spins. 
This implies also that the final results of shell model calculations 
do not depend on the choice of
the single-particle states (spherical or deformed) in 
building a shell model space. 

Fig. 1 indicates that the PSM is a reasonable shell model
truncation scheme as it reproduces the result of the large-scale pf-SM very well.
Furthermore, the advantage of the PSM is that it is able to extract the 
undergoing physics. Here, it is to understand why and how the
backbending occurs. A band diagram displays band
energies of various configurations before they are mixed by the
diagonalization procedure Eq.(\ref{psmeq}). 
Irregularity in a spectrum may appear if a
band is crossed by another one at certain spin. 
It was found \cite{Cr48}
that in $^{48}$Cr, the backbending is caused by a band crossing
that corresponds to a simultaneous alignment of the $f_{7/2}$
neutron and proton pairs. 

Following the success in the $^{48}$Cr calculation, it has been  
demonstrated in Ref. \cite{Odd50} that the PSM can, using the same
set of parameters as used for $^{48}$Cr, 
describe the odd-mass nuclei $^{47,49}$V,$^{47,49}$Cr
and $^{49,51}$Mn, which have also been extensively studied with the large-scale 
pf-SM
diagonalizations \cite{Pin97}. 
Another recent PSM work \cite{psmoo} 
has shown that the model can be applied to
the odd-odd nuclei near the N = Z line as well.

\section{The N $\approx$ Z nuclei in the mass-80 region}
\label{sec:2}

\subsection{Properties of the yrast bands}

In N = Z nuclei, neutrons and protons occupy the same 
orbitals, and thus can have the largest probability to interact with each other.
The study of N = Z nuclei is the domain which is expected
to give the most relevant information about the properties of
the neutron-proton (n-p) interaction.
An interesting question for the yrast band properties is whether 
the expected strong n-p interaction will modify the rotational-alignment picture
in N = Z nuclei. 
It has been suggested using the cranking
approaches in a single-$j$ shell (see the references cited in Ref. \cite{SuSh}) 
that the rotational-alignment
properties can be modified by the residual n-p
interaction. 
The recently observed alignment delays in $^{72}$Kr, $^{76}$Sr,
and $^{80}$Zr \cite{Kr72} were 
conjectured as possible consequences of the strong 
n-p pairing interaction. 

Experimentally locating the band crossings 
in these N $\approx$ Z nuclei 
is a very challenging task. 
The main difficulty in extending the study of N = Z nuclei
to high spins is that their population has extremely low cross sections in
the small number of available reactions.
Despite of the difficulties, progress in the
development of large $\gamma$-ray arrays and associated ancillary detectors,
and refinements of the data processing techniques,
has allowed recent advance in the
knowledge of some heavier N = Z nuclei (see, for example, Refs. 
\cite{Kr72,Mo84,Ru88,Kr72Ber}).

Parallel to the experimental efforts,  
a systematic investigation for the yrast properties of the
N = Z, Z+2, and Z+4 in Kr, Sr and Zr nuclei has been carried out by
the PSM \cite{psmee}.
The calculations described reasonably well  
most of the observables (moments
of inertia, transition quadrupole moments and g-factors) 
known for these nuclei, with a notable exception: the
very pronounced delay in the crossing frequency in $^{72}$Kr \cite{Kr72}. 
To understand this disagreement, 
a more detailed PSM analysis for the N = Z Kr, Sr and Zr nuclei was 
subsequently performed \cite{SuSh}.
While inclusion of an empirically enhanced residual
n-p interaction in the Q-Q channel
was found helpful to improve the agreement with the data, it was realized also
that this enhancement offsets
the good agreement previously obtained for the N $\ne$ Z nuclei $^{74,76}$Kr
with the standard interaction strength \cite{SuSh}
(see Fig. 2 below).

Very recently, the observation of another band in $^{72}$Kr has been reported  
\cite{Kr72Ber}, which solved our puzzle why the $^{72}$Kr rotational-alignment
should exhibit such an unusual behavior as suggested in Ref. \cite{Kr72}. 
Calculations show that this new band could be 
the missing aligned S-band, which lies
much closer to what was predicted \cite{psmee}. 

% For tables use
\begin{table}
\caption{Deformations $\varepsilon_2$ 
at which the projected bases are constructed for the nuclei 
presented in Fig. 2. }
\label{tab:2}       % Give a unique label
% For LaTeX tables use
\begin{tabular}{cccc}
\hline\noalign{\smallskip}
N = Z nuclei & $\varepsilon_2$ & N = Z+2 nuclei & $\varepsilon_2$ \\ 
\noalign{\smallskip}\hline\noalign{\smallskip}
$^{72}$Kr & 0.36 & $^{74}$Kr & 0.36 \\
$^{76}$Sr & 0.36 & $^{78}$Sr & 0.36 \\
$^{80}$Zr & 0.36 & $^{82}$Zr & 0.29 \\
$^{84}$Mo & 0.28 & $^{86}$Mo & 0.22 \\
$^{88}$Ru & 0.23 & $^{90}$Ru & 0.16 \\
\noalign{\smallskip}\hline
\end{tabular}
% Or use
%\vspace*{5cm}  % with the correct table height
\end{table}
\begin{figure}
\resizebox{0.47\textwidth}{!}{%
  \includegraphics{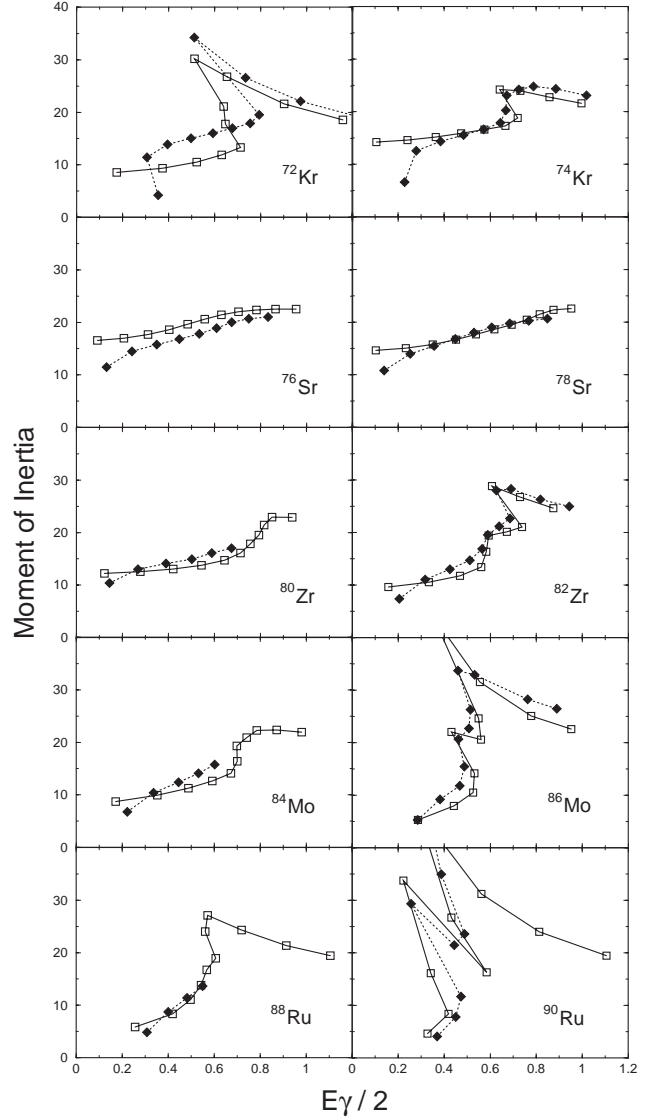}
}
\caption{Comparison of the calculated yrast energy levels with known data
for $N=Z$ and $N=Z+2$ nuclei in the plots of moment of inertia
$\Im^{(1)}(I) = (2I-1)/E_\gamma(I\rightarrow I-2)$ vs.
rotational frequency $\omega(I) = E_\gamma(I\rightarrow I-2)/2$.}
\label{fig:1}       % Give a unique label
\end{figure}

To show to what extent the yrast properties of these nuclei can be
understood by the PSM in a systematical way, we have 
calculated the N = Z and N = Z+2 isotopes
from Z = 36 to 44. The calculations are performed by using three major shells
($N=2,3,4$) for both neutrons and protons, with the pairing
interaction strengths 
$G_1=20.25$ and $G_2=16.20$, and the quadrupole-pairing constant
$G_Q=0.16 G_M$, 
for all the nuclei considered (These strengths are the same as those used 
in Ref. \cite{psmee}). 
Table 1 lists the deformation parameters, at which the shell model spaces
are constructed. 
The results are presented in Fig. 2.
 
As can be seen, the standard PSM calculations describe quite well the 
moments of inertia of the N = Z+2 nuclei up to the highest measured spin.
In particular, the fine structure at the band crossing regions is
quantitatively reproduced.  
The calculations also predict well
the known data for the N = Z nuclei in this region.
Going from lighter to heavier nuclei 
along the N = Z line, one finds that the degree of
band disturbance varies, exhibiting 
back-bend (Kr), no bend (Sr), slight up-bend (Zr),
up-bend (Mo), back-bend (Ru).  
The PSM analysis suggests that this 
systimatical behavior in band
disturbance may be analogous to what has been known in the rare-earth
region \cite{Hama}. 
From Kr to Ru isotopes, the closest orbitals to the Fermi
levels move gradually from $K=1/2$ and 3/2 to $K=5/2$ and 7/2 states of 
the $g_{9/2}$ shell.  
Since the crossing bands are the 2-qp states having components from
the near-Fermi orbitals, different shell fillings make the effective band
interactions at the crossing region 
rather different, resulting in different disturbances in the yrast bands. 

However, 
it is still an open question whether the standard set of the PSM interactions is
sufficient to describe the yrast properties of these N = Z nuclei. 
The present data for $^{80}$Zr \cite{Kr72}, $^{84}$Mo \cite{Mo84}, and
$^{88}$Ru \cite{Ru88} stop just before the spin states where band disturbances
are predicted to occur. 
The PSM calculations with an empirically enhanced n-p interaction 
(results not shown) 
suggested higher
alignment frequencies \cite{Mo84}. 
Thus, although the existing experimental data seem to indicate some delays 
of the alignment frequency in comparison with the neighboring $N=Z+2$
nuclei, one cannot immediately relate 
this to an enhancement of the n-p interaction.
From Fig. 2 one may only conclude 
that at least two or three more yrast transitions 
are very much
desired in these $N=Z$ nuclei for a better testing of the PSM predictions. 
  
\subsection{$K$-isomeric states}

In an axially-deformed system, $K$ is a good quantum number. 
Conservation of $K$ leads to selection rules for transitions involving 
changes in $K$ between the initial and final states, which in turn
lead to decay hidrances resulting in isomers.
Depending on the degree of the transition hindrance, 
some high-$K$ isomeric states
can be very long-lived \cite{Walker}.
Long-lived
isomers are not only interesting from the
structure point of view, they may also be found important  
in the interdisciplinary fields.
For nuclear astrophysics, information regarding the structure of
isomeric states and the nearby states is very valuable. 
One recent example \cite{Ta180,Ta180W}
is the study of the excitation and decay of $^{180}$Ta$^m$
in understanding the nuclear astrophysics puzzle on the production and
survival of $^{180}$Ta in stars.  

It has been argued  
that the 
existence of isomers in nuclei along the rp-process path 
could significantly modify the current conclusions
on the nucleosynthesis and the correlated energy generation
in X-ray bursts \cite{SchatzRep}.
Along the N = Z line, there are waiting point nuclei 
whose lifetimes entirely determine the speed of nucleosynthesis towards
heavier nuclei and the produced isotopic abundances. The lifetimes are 
strongly dependent on the photodisintegration rates,  
the $\beta$-decay rates, and particularly on the masses
of nuclei along the path.
If there are isomeric states in the low-energy region in
these nuclei,
the total $\beta$-decay half-life of an isotope, 
for example, can be affected strongly
if an isomeric state is populated and has a significantly different
$\beta$-decay half-life compared to the ground state. 
Therefore, knowing the existence and the
structure of the isomers could have
a significant impact. 

Shape co-existence phenomenon has been observed in the
mass-80 region.
There is evidence that at Z $\sim$ 34 
the oblate shape can be energetically favored, while for heavier nuclei
(with Z $\sim$ 44) the prolate shape dominates. 
It is interesting to realize that near the N $\approx$ Z line, in both deformed 
energy minima (oblate for Z $\sim$ 34 and prolate for Z $\sim$ 44 systems),
the $K$-components of the $g_{9/2}$ sub-shell are found to be high.
Among the nuclei discussed
in this paper, the $K=7/2$ and 9/2 orbitals lie close to
the Fermi levels in the Se isotopes at oblate deformations, and 
the $K=5/2$ and 7/2 orbitals are found near  
the Fermi levels in the
Zr, Mo, and Ru isotopes at their prolate minima. 
Quasi-particles of these orbitals can couple
to a $K=8$ 2-qp state in the former, 
and a $K=6$ 2-qp state in the latter case. 

% For tables use
\begin{table}
\caption{The predicted energy levels of $I^\pi = 6^+$ isomeric state
(relative to the ground state) in $N=Z$ and
$N=Z+2$ nuclei. The calculated $I^\pi = 6^+$ levels in the ground band
are also shown. }
\label{tab:1}       % Give a unique label
% For LaTeX tables use
\begin{tabular}{lcc}
\hline\noalign{\smallskip}
Nuclei & E($I^\pi = 6^+_{\rm ground}$) (MeV) & 
E($I^\pi = 6^+_{\rm isomer}$) (MeV) \\
\noalign{\smallskip}\hline\noalign{\smallskip}
$^{80}$Zr & 1.65 & 4.19 \\
$^{82}$Zr & 1.91 & 3.71 \\
$^{84}$Mo & 2.03 & 3.53 \\
$^{86}$Mo & 2.51 & 3.21 \\
$^{88}$Ru & 2.35 & 3.14 \\
$^{90}$Ru & 2.17 & 2.38 \\
\noalign{\smallskip}\hline
\end{tabular}
% Or use
%\vspace*{5cm}  % with the correct table height
\end{table}

In Table 2, the predicted energy levels of the $I^\pi = 6^+$ isomeric states
(relative to the ground state) in the $N=Z$ and
$N=Z+2$ Zr, Mo, and Ru isotopes are given. 
Our prediction is based on realistic
calculations that have reasonably reproduced all the known structure data
in these nuclei (see Fig. 2).
To indicate how high these states lie above the yrast band, energies of the
$I^\pi = 6^+$ levels in the ground band are also shown.
It is worth pointing out that 
the same $K$-components that couple to a $K=6$ 2-qp state
can give rise to a crossing band if they couple to a $K=1$ 2-qp
state, causing the band disturbances as discussed in Fig. 2. 

\subsection{Structure of the waiting-point nucleus $^{68}$Se}

The predicted 
shape coexistence in $^{68}$Se seems to be firmly suggested
by the recent experimental data \cite{Se68}.
Two coexisting rotational bands were identified, with the ground state band
having properties consistent with collective oblate rotation, and the
excited band having characteristics consistent with prolate rotation.
Our PSM calculations are performed for $^{68}$Se at deformation 
$\varepsilon_2=0.28$ for the prolate states,
and $\varepsilon_2=-0.24$ for the oblate states.
The calculation conditions are the same as used for the nuclei in Fig. 2,  
except for slightly reduced pairing strengths: $G_1=19.58$ and $G_2=15.66$. 
In addition, 
we found that the recently suggested Nilsson parameter set \cite{Sun00}
works better for the description of $^{68}$Se.

Fig. 3 compares the results of $^{68}$Se for the
lowest levels of each spin, and at the prolate and oblate
minima. 
Good agreement has been obtained for the known oblate
band. The calculations suggest a gradual increase in the moment of inertia
extended to higher spin states beyond the current experimental band \cite{Se68}.
In contrast, the moment of inertia of the prolate band behaves rather 
differently. The data bend back sharply as the rotational frequency increases. 
This basic feature in the
prolate band has also been reproduced by the calculation.
In addition to the observed sharp backbending at spin 8,
we predict a second sharp backbending at spin 16. 

\begin{figure}
\resizebox{0.42\textwidth}{!}{%
  \includegraphics{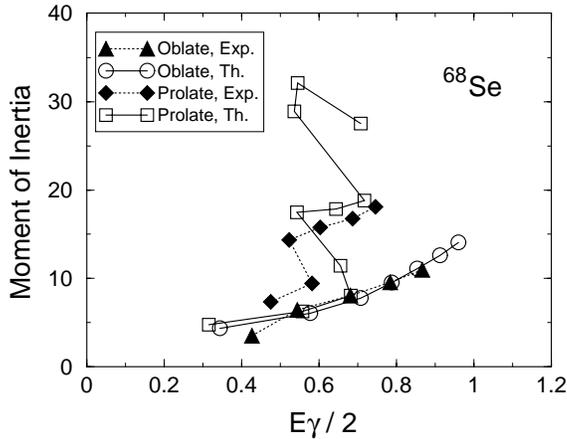}
}
\caption{Comparison of the calculated yrast energy levels with known data
in $^{68}$Se in the plot of Moments of inertia 
$\Im^{(1)}(I) = (2I-1)/E_\gamma(I\rightarrow I-2)$ vs. rotational frequency   
$\omega(I) = E_\gamma(I\rightarrow I-2)/2$. 
The experimental data are taken from Ref.    
\protect\cite{Se68}. 
}
\label{fig:3}       % Give a unique label
\end{figure}

At the prolate minimum in $^{68}$Se, 
the Fermi levels are surrounded by the low-$K$ states,
whereas at the oblate minimum, they are near the high-$K$ states.
The low- and high-$K$ states respond rather differently to the rotation,
which is reflected in very different rotational alignments.
Detailed analysis \cite{Se68theory} using the band diagrams suggested 
that the  $K=1$ band based on $g_{9/2}$ quasiparticles with $K=7/2$ and 
$9/2$ at the oblate minimum crosses the 0-qp band so gently that 
one cannot see a band disturbance
in the yrast solution.
However, the  $K=1$ band based on $g_{9/2}$ quasiparticles with $K=1/2$ and 
$3/2$ at the prolate minimum crosses the 0-qp band sharply. In the latter case,
the regular band is disturbed, resulting in
the observed first bankbending. 
The predicted second backbending 
is caused by a crossing of the 4-qp band that consists of a proton and a neutron
pair \cite{Se68theory}.  

In the same calculation, 
we predict 
high-$K$ bands ($K=8$) with a bandhead of 
spin 8 and an excitation energy of 5 MeV. They are
long-lived isomeric states in the sense that no allowed $\gamma$-transition
matrix elements of low multipolarity
can connect these states to the nearby ground band ($K=0$).
Thus, once the isomeric states are populated,
the high-$K$ states find no path for further $\gamma$-decay.
An analysis with inclusion of the isomeric states in 
the rp-process network simulations is in progress \cite{Se68net}. 

\section{Summary}
\label{sec:4}

One of the frontiers in nuclear physics is the study of nuclei far from
the $\beta$-stability. It is desired to apply advanced shell model
diagonalization methods while pushing the calculations to heavy nuclear
systems. However, most of the spherical shell model calculations are
severely limited in the fp-shell nuclei or nuclei in the vicinity of
shell closures.

We have shown that the projected shell model may be an efficient
truncation scheme of the exact shell model solution. 
In the calculation for $^{48}$Cr, the PSM results are comparable with 
those obtained by
the large-scale shell model calculations.  
Systematic calculations
are performed for the collective levels
as well as the quasi-particle excitations for the N $\approx$ Z nuclei 
in the mass-80 region.
While the known experimental data
of the yrast bands (from Se to Ru)
are quantitatively
described, the present
calculations predict the existence of $K$-isomeric states,
some of which lie low in energy under certain structure conditions.
We hope that these states can be identified experimentally.
It will be highly interesting to see how much their presence modifies the
current conclusions of the rp-process nucleosynthesis.

The author wishes to express his sincere thanks to D. Bucurescu and C.A. Ur
for motivating him to study 
this interesting mass region. 
Valuable discussions with J.A. Sheikh and C.J. Lister, as well as with
A.V. Afanasjev, A. Aprahamian, S. Frauendorf, and M. Wiescher are acknowledged. 
This work is supported by the NSF under contract number PHY-0140324.

%
% For two-column wide figures use
%\begin{figure*}
% Use the relevant command for your figure-insertion program
% to insert the figure file. See example above.
% If not, use
%\vspace*{5cm}       % Give the correct figure height in cm
%\caption{Please write your figure caption here}
%\label{fig:2}       % Give a unique label
%\end{figure*}
%
%
% BibTeX users please use
% \bibliographystyle{}
% \bibliography{}

\begin{thebibliography}{}
%
% and use \bibitem to create references.
%
\bibitem{SchatzRep} H. Schatz, {\it et al.}, Phys. Rep. \textbf {294} (1998) 167.

\bibitem{SchatzPRL} H. Schatz, {\it et al.}, Phys. Rev. Lett. \textbf {86} (2001) 
3471.

\bibitem{psm}
K.~Hara and Y.~Sun, Int. J. Mod. Phys. \textbf {E4} (1995) 637.

\bibitem{MV} K.W. Schmid and F. Gr\"ummer, Rep. Prog. Phys.
\textbf {50}, 731 (1987).

\bibitem{FDSM} C.\,-L.\ Wu,  D.\ H.\ Feng and M.\ Guidry,
Adv.\ Nucl.\ Phys.\ \textbf {21} (1994) 227. 

\bibitem{PSsu3} G. Popa, J.G. Hirsch and J.P. Draayer, Phys. Rev. 
\textbf {C62} (2000) 064313.

\bibitem{MC} T. Otsuka, Nucl. Phys. \textbf {A704} (2002) 21c. 

\bibitem{RS.80}
 P. Ring and P. Schuck, \textit {The Nuclear Many Body Problem}
 (Springer-Verlag, New York, 1980). 

\bibitem{fpshell}
E. Caurier, {\em et al}, Phys. Rev. Lett. \textbf {75} (1995) 2466.

\bibitem{HI80} K. Hara and S. Iwasaki, Nucl. Phys. \textbf {A332} (1979) 61; 
\textbf {A348} (1980) 200.

\bibitem{code} Y.~Sun and K.~Hara, Comput. Phys. Comm. \textbf {104} (1997) 245.

\bibitem{brkm}
T. Bengtsson and I. Ragnarsson, Nucl. Phys. \textbf {A436} (1985) 14. 

\bibitem{Zuker96}
M. Dufour and A.P. Zuker, Phys. Rev. \textbf {C54} (1996) 1641. 

\bibitem{pairing}
W. Dieterich, {\em et al}, Nucl. Phys. \textbf {A253} (1975) 429. 

\bibitem{Cr48} K. Hara, Y. Sun and T. Mizusaki,
 Phys. Rev. Lett. \textbf {83} (1999) 1922.

\bibitem{exp_b}
F. Brandolini, {\em et al}, Nucl. Phys. \textbf {A642} (1998) 387. 

\bibitem{Odd50} 
V. Vel\'azquez, J.G. Hirsch and Y. Sun, Nucl. Phys. \textbf {A686} (2001) 129.

\bibitem{Pin97} G. Mart\'\i nez-Pinedo, A.P. Zuker, A. Poves and
E. Caurier, Phys. Rev. \textbf {C55} (1997) 187.

\bibitem{psmoo} R.~Palit, J.A.~Sheikh, Y.~Sun, and H.C.~Jain, Phys. Rev. 
\textbf {C67} (2003) 014321.   

\bibitem{SuSh} Y. Sun and J.A.~Sheikh, Phys. Rev. \textbf {C64} (2001) 
031302(R).

\bibitem{Kr72} S.M. Fischer, {\it et al.}, Phys. Rev. Lett. \textbf {87}
(2001) 132501.

\bibitem{Mo84} N.~M\u arginean, {\it et al.}, Phys. Rev. \textbf {C65} 
(2002) 051303(R). 

\bibitem{Ru88} N.~M\u arginean, {\it et al.}, Phys. Rev. \textbf {C63} 
(2001) 031303(R). 

\bibitem{Kr72Ber} N.S. Kelsall, {\it et al.}, Proceedings of the Conference
on Frontiers of Nuclear Structure, Berkeley, USA, Jul. 29 - Aug. 2, 2002.

\bibitem{psmee} R.~Palit, J.A.~Sheikh, Y.~Sun, and H.C.~Jain,
Nucl. Phys. \textbf {A686} (2001) 141.

\bibitem{Hama} R. Bengtsson, I. Hamamoto and B.R. Mottelson, Phys. Lett.
\textbf {73B} (1978) 259. 

\bibitem{Walker} P. Walker and G. Dracoulis, Nature (London) 
\textbf {399} (1999) 35.

\bibitem{Ta180} D. Belic, {\it et al.}, Phys. Rev. Lett. \textbf {83}
(1999) 5242; Phys. Rev. \textbf {C65} (2002) 035801.

\bibitem{Ta180W} P.M. Walker, G.D. Dracoulis and J.J. Carroll,
Phys. Rev. \textbf {C64} (2001) 061302(R).

\bibitem{Se68} S.M. Fischer, {\it et al.}, Phys. Rev. Lett. \textbf {84}
(2000) 4064.

\bibitem{Sun00} Y. Sun, {\it et al.},
 Phys. Rev. \textbf {C62} (2000) 021601(R).

\bibitem{Se68theory} Y. Sun, Z.-W. Ma, A. Aprahamian and M. Wiescher,
nucl-th/0107004. 

\bibitem{Se68net} A. Aprahamian and M. Wiescher, private communication.

\end{thebibliography}
%
% Non-BibTeX users please use

\end{document}